# ATTENTIVE ITEM2VEC: NEURAL ATTENTIVE USER REPRESENTATIONS

*Oren Barkan\*[1], Avi Caciularu\*[12], Ori Katz[13] and Noam Koenigstein[14]*

[1]Microsoft
[2]Bar-Ilan University
[3]Technion
[4]Tel Aviv University

## ABSTRACT

Factorization methods for recommender systems tend to represent users as a single latent vector. However, user behavior and interests may change in the context of the recommendations that are presented to the user. For example, in the case of movie recommendations, it is usually true that earlier user data is less informative than more recent data. However, it is possible that a certain early movie may become suddenly more relevant in the presence of a popular sequel movie. This is just a single example of a variety of possible dynamically altering user interests in the presence of a potential new recommendation. In this work, we present Attentive Item2vec (AI2V) - a novel attentive version of Item2vec (I2V). AI2V employs a context-target attention mechanism in order to learn and capture different characteristics of user historical behavior (context) with respect to a potential recommended item (target). The attentive context-target mechanism enables a final neural attentive user representation. We demonstrate the effectiveness of AI2V on several datasets, where it is shown to outperform other baselines.

## 1. INTRODUCTION AND RELATED WORK

Item2vec (I2V) [10] is an increasingly popular embedding model for learning item representations based on Collaborative Filtering (CF) data. This work builds upon I2V and presents Attentive Item2vec (AI2V) – a model that learns neural attentive user representations that dynamically change in the presence of different potential item recommendations.

Traditional CF methods [6], [7], such as Matrix Factorization models (MF) represent users as a single user vector that remains static regardless of the potential recommendation being considered. However, user interests and behavior form a context that may dynamically change in the presence of different recommended target items. Consider the following example: A user may be generally interested in popular main-stream action movies and less interested in niche less popular action movies. That user may also have a minor interest in a specific historical era, e.g., the Roman Empire or the US Civil War and as a result, she has watched some documentaries on that subject a while back. These documentaries are generally not very relevant to her current interests, however in the presence of a niche (not very popular) movie about the Roman Empire, the "old" events become much more relevant.

A key contribution of AI2V arises from its novel neural attentive user representation that dynamically adjusts in the presence of different target item recommendations: first, AI2V extracts attentive context-target representations each designed to capture different user behavior w.r.t the potential target item recommendation. These attentive context-target representations are then fed into a secondary network that chooses the most relevant user characteristics and constructs a final attentive user vector. This vector constitutes the dynamic representation of that user w.r.t. the specific target item recommendation.

Attention in neural networks are investigated in the context of various applications such as recommendations [13], [16], decoding error correction codes, [25], and NLP tasks [14]. In the latter, early attention models employed recurrent mechanisms [15]. However, a recent trend in the NLP community is to utilize attention mechanisms directly on the input embeddings [17] instead of RNNs. This setup was successfully employed by Vaswani et al. [18], where it was shown to produce superior results on multiple machine translation tasks.

AI2V employs a novel context-target attention mechanism to form a final neural attentive user representation that is subsequently used for scoring the similarity between the user and the potential target item recommendation. Different from [18], AI2V does not employ self-attention but context-target attention (where the context is the user's historical items and

---

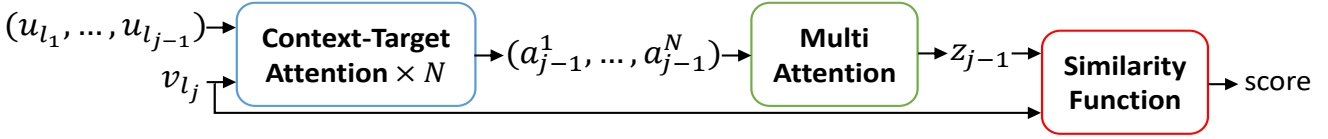

Figure 1. The AI2V model. The context vectors and the target vector that correspond to the context items in the sub-user $x_{1:j-1}$ and the target item $l_j$, respectively, are input to the $N$ independent context-target attention units (blue) that output $N$ attentive context-target vectors $\{a_{j-1}^i\}_{i=1}^N$ (Eq. (5)). The attentive context-target vectors are input to the multi attention module (green) that outputs the final neural attentive user vector $z_{j-1}$ (Eq. (5)). Finally, the similarity between $z_{j-1}$ and $v_{l_j}$ is computed (red) using Eq. (7).

the target is the potential item recommendation). Also, AI2V scores the attention between the target item to the context of historical items by the cosine similarity function instead of the scaled dot-product [18]. Lastly, AI2V learns a neural scoring function for scoring the similarity between the attentive user and item vectors.

Further attempts for incorporating attention mechanisms in recommender systems include [4] and [5]. However, these models leverage auxiliary content-based information [23] about the items and hence can be seen as hybrid models rather a pure CF approach such as AI2V.

The remainder of the paper is organized as follows: Section 2 briefly describes I2V, which is a related work. Section 3 presents the AI2V model itself and is followed by evaluations in Section 4 and a summary in Section 5.

## 2. ITEM2VEC (I2V)

In this section, we provide a brief overview of the I2V model [10], which serves as the basis for AI2V. Let $I = \{i\}_{i=1}^J$ be a set of all item identifiers. For each item $i$, I2V learns latent context and target vectors $u_i, v_i \in \mathbb{R}^d$. These latent vectors are estimated via implicit factorization of items co-occurrences matrix. Specifically, we represent a user $x = (l_1, \ldots, l_K)$ as a list of historical items that were co-consumed together. We consider, W.l.o.g, a dataset of a single user $x$ (where the extension to multiple users is straightforward). The I2V objective is to minimize the loss function

$$L_x = -\sum_{i=1}^{K}\sum_{j\neq i}^{K} \log p(l_j|l_i) \quad (1)$$

with

$$p(l_j|l_i) = \sigma(s(l_i,l_j)) \prod_{k \in \mathcal{N}} \sigma(-s(l_i,k)) \quad (2)$$

where $s(i,j) = u_i^T v_j$, $\sigma(x) = (1 + \exp(-x))^{-1}$ and $\mathcal{N} \subset I$ is a subset of items that are sampled from $I$ according to the unigram item distribution raised to the 0.5 power [24]. The items in $\mathcal{N}$ are treated as negative target items w.r.t. the context item $l_i$. In order to mitigate the popularity bias in common CF datasets, I2V further applies a subsampling procedure in which items are randomly discarded from the users according to their popularity. The amount of subsampling is controlled by a hyperparameter that is adjusted w.r.t the dataset (we refer the reader to [10] for further details).

In the training phase, I2V learns the sets of context and target vectors $U, V \subset \mathbb{R}^d$ by minimizing $L_x$ using stochastic gradient descent. In the inference phase, the similarity between context and target item vectors $u_i$ and $v_j$ is computed via the cosine similarity

$$h(u_i, v_j) = \frac{u_i^T v_j}{|u_i||u_j|}. \quad (3)$$

## 3. ATTENTIVE ITEM2VEC (AI2V)

AI2V differs from I2V in several aspects: First, AI2V models the probability of the target item $l_j$ given all previous items in the user history $l_{1:j-1}$ and hence optimizes a different objective than Eq. (1). Secondly, AI2V employs softmax with negative sampling, instead of the skip-gram with negative sampling [11] that appears in Eq. (2). Third, AI2V employs a novel attention mechanism that learns to attend the context items within the user history according to their consumption order and w.r.t to the target item. Fourth, AI2V models the similarity between the attentive user and candidate item vectors using a neural scoring function, which is different from the dot-product function used by I2V. Lastly, AI2V introduces target biases, which eliminate the need for the subsampling procedure from [10], which requires an additional hyperparameter tuning effort.

### 3.1. Attentive Context-Target Representation

In this section, we describe the attention mechanism employed by AI2V. Consider again a user $x = (l_1, \ldots, l_K)$ and denote a sub-user by $x_{1:j-1} = (l_1, \ldots, l_{j-1})$ $(j-1 < K)$. AI2V models the attentive context-target representation of $x_{1:j-1}$ via

$$a_{j-1} = \sum_{m=1}^{j-1} \alpha_{jm} B_c u_{l_m}, \quad (4)$$

where $B_c \in \mathbb{R}^{d \times d}$ is a linear mapping that maps the historical context item vectors to a new space and $\alpha_{jm}$ are the attention weights that are computed as

$$\alpha_{jm} = \frac{\exp\left(h(A_c u_{l_m}, A_t v_{l_j})\right)}{\sum_{n=1}^{j-1} \exp\left(h(A_c u_{l_n}, A_t v_{l_j})\right)}$$

where $A_c, A_t \in \mathbb{R}^{d_a \times d}$ are learnable linear mappings from the context and target spaces to $d_a$-dimensional context and target attention spaces, respectively. $h: \mathbb{R}^{d_a} \times \mathbb{R}^{d_a} \to \mathbb{R}$, is the cosine similarity function defined in Eq. (3). The context-target attention module appears in Fig. 1 (blue).

### 3.2. Neural Attentive User Representation

We propose learning of $N$ context-target attention mechanisms (Section 3.1) in parallel. Each mechanism is associated with a different set of learnable parameters and functions $\{A_c^i, A_t^i, B_c^i\}_{i=1}^{N}$ that produces the attentive context-target representations $\{a_{j-1}^i\}_{i=1}^{N}$ using Eq. (4). Then, the multi attentive user representation is computed as

$$z_{j-1} = R w_{j-1} \tag{5}$$

where $w_{j-1} = \left[(a_{j-1}^1)^T, \ldots, (a_{j-1}^N)^T\right]^T$ and $R \in \mathbb{R}^{d \times Nd}$ is a learnable linear mapping. This allows AI2V to learn various types of attention functions and aggregate the information that is extracted by each attention function into the final multi attentive user representation (marked green in Fig. 1).

### 3.3. AI2V Similarity Function

The neural attentive user representation $z_{j-1}$ from Eq. (5) encodes the several attentive context-target representations. We use it together with the item representation as an input to our scoring function. AI2V replaces the dot-product scoring function $s$ from Eq. (2) with the following neural scoring function $\psi: \mathbb{R}^d \times \mathbb{R}^d \to \mathbb{R}$,

$$\psi(u, v) = W_1 \phi(W_0([u, v, u \circ v, |u - v|])), \tag{6}$$

where $[\cdot]$ denotes concatenation operator, $\circ$ denotes the Hadamard product, $\phi$ is the ReLU activation function, and $W_0 \in \mathbb{R}^{d \times 4d}$ and $W_1 \in \mathbb{R}^{1 \times d}$ are learnable linear mappings (matrices). This is a neural network with a single ReLU activated hidden layer and a scalar output. This scoring function (which is inspired by [22]) was shown to improve results on several natural language understanding tasks comparing to a simple scoring such as dot product.

The final similarity score for $(x_{1:j-1}, l_j)$ is computed by

$$o\left(z_{j-1}, v_{l_j}\right) = \psi_o\left(z_{j-1}, B_t v_{l_j}\right) + b_{l_j} \tag{7}$$

where $\psi_o$ is a neural scoring function in the form of Eq. (4), $B_c \in \mathbb{R}^{d \times d}$, and $b_{l_j}$ is a target bias term that corresponds to the item $l_j$. The similarity component is illustrated in red as part of the whole pipeline that appears in Fig. 1.

### 3.4. AI2V Training and Inference

Given a user $x = (l_1, \ldots, l_K)$. AI2V objective is to minimize the loss function

$$L_x = \sum_{j=2}^{K} -\log p(l_j | l_{1:j-1}), \tag{8}$$

with

$$p(l_j | l_{1:j-1}) = \frac{\exp\left(o\left(z_{j-1}, v_{l_j}\right)\right)}{\sum_{k \in \mathcal{N}} \exp\left(o(z_{j-1}, v_k)\right)}.$$

Note that the AI2V and I2V objectives and scoring functions are different: Eq. (2) vs. Eq. (8), and $s$ (Eq. (2)) vs. $\psi$ (Eq. (6)). Finally, in the inference phase, AI2V uses the same similarity function $o$ from Eq. (7) that is used during the training phase, whereas I2V uses the cosine similarity according to Eq. (3).

## 4. EXPERIMENTAL SETUP AND RESULTS

In this section, we present the evaluation measures and protocol, datasets and data preparation, evaluation methods, hyperparameters configuration, optimization details and results.

### 4.1. Evaluation Measures and Protocol

The evaluation protocol is as follows: for each user $x = (l_1, \ldots, l_K)$, we create $K - 1$ examples $\{(x_{1:j-1}, l_j)\}_{j=2}^{K}$ where the task is to predict the target item $l_j$ for the query sub-user $x_{1:j-1} = (l_1, \ldots, l_{j-1})$. We use two popular [1]-[3], [12], [16] evaluation measures in our experiments:

**Hit Ratio at $K$ (HR@$K$)**: This measure is defined as the percentage of the predictions made by the model, where the true item was found in the top K items suggested by the model. Specifically, a query-target pair $(x_{1:j-1}, l_j)$, is scored with 1 if the target item $l_j$ is ranked in the top $K$ recommendations produced by the model w.r.t. to the query $x_{1:j-1}$ otherwise 0. Then we average over all query-target examples in the test set. Note that this measure ignores the order of the recommended items at the top.

**Mean Reciprocal Rank at $K$ (MRR@$K$)**: This measure is defined as the average of the reciprocal ranks [1], where the reciprocal rank is set to zero if the rank is above $K$. Unlike HR@$K$, the MRR@$K$ measure does consider the order of the recommendation list.

### 4.2. Datasets and Data Preparation

Three datasets are considered, which contain historical items that were consumed by the user. Data preparations were performed as follows:

**MS**: This is a proprietary dataset from Microsoft Store that contains 4 months of historical user purchases (ordered by time). Users with less than 4 purchases were filtered, and the first 3 months of the data were used for training the model. The last month was used for testing. For example, consider a user $x = (l_1, \ldots, l_K)$ that purchased the items in $x_{j:K}$ during the last month. Then, the test examples that are associated with this user are $\{(x_{1:m-1}, l_m)\}_{m=j}^{K}$ and the training examples are $\{(x_{1:m-1}, l_m)\}_{m=2}^{j-1}$.

Table 1. Datasets statistics

| Dataset | #items | #users | #training examples | #test examples |
|---|---|---|---|---|
| MovieLens | 6,040 | 3,577 | 778,679 | 6,040 |
| MS | 2,138 | 10,000 | 95,753 | 10,000 |
| Yahoo! | 4,222 | 9,362 | 241,896 | 9,362 |

Table 2. Performance Evaluation on MS dataset ($\cdot\ 10^{-2}$)

| Model | HR@5 | HR@10 | HR@20 | MRR@5 | MRR@10 | MRR@20 |
|---|---|---|---|---|---|---|
| **AI2V** | **20.81** | **26.95** | **34.89** | **12.88** | **13.70** | **14.24** |
| I2V | 11.06 | 15.53 | 22.37 | 4.53 | 5.13 | 5.59 |
| NCF | 11.42 | 20.08 | 32.28 | 5.59 | 6.70 | 7.55 |

Table 3. Performance Evaluation on MovieLens dataset ($\cdot\ 10^{-3}$)

| Model | HR@5 | HR@10 | HR@20 | MRR@5 | MRR@10 | MRR@20 |
|---|---|---|---|---|---|---|
| **AI2V** | **12.92** | **28.06** | **58.56** | **5.56** | **7.39** | **9.52** |
| I2V | 11.30 | 27.05 | 55.73 | 3.75 | 5.87 | 7.75 |
| NCF | 8.48 | 19.99 | 45.84 | 3.25 | 4.78 | 6.56 |

Table 4. Performance Evaluation on Yahoo! Music dataset ($\cdot\ 10^{-1}$)

| Model | HR@5 | HR@10 | HR@20 | MRR@5 | MRR@10 | MRR@20 |
|---|---|---|---|---|---|---|
| **AI2V** | **8.42** | **14.81** | **24.98** | **3.93** | **4.77** | **5.46** |
| I2V | 5.65 | 10.94 | 18.07 | 2.57 | 3.26 | 3.75 |
| NCF | 5.64 | 11.09 | 19.41 | 2.57 | 3.28 | 3.85 |

**MovieLens**: This dataset is based on the MovieLens 1M dataset [20] containing movie ratings using a 5-star scale (0.5 - 5.0). From each user's rating list, we considered all the movies in the sequence with ratings above 3.5. We further filtered all users with less than two items. The split for training and test examples is done by taking the last movie ranked by each user to be the test set (according to the MovieLens timestamps), and the training data consists of the rest of the data.

**Yahoo! Music**: This is a Yahoo dataset [21] based on user ratings on a 100-score integer scale (0 – 100), with a special score of 255 given to songs that were extremely disliked. The preparation of the data was done in the same manner as done with the MS data, taking into consideration songs that were ranked above the score of 80 as positive examples. Items rated 255 and below 80 were not considered as positive examples for that user.

In order to speed up training, we used a sparse version of the above datasets - several users and items from the catalog were randomly sampled. Furthermore, we removed the 20 most popular items from the test set, according to the popularity determined by the training set. These items are promoted automatically and are less interesting as a recommendation. The datasets' statistics are summarized in Table 1.

### 4.3. Evaluated Methods

The evaluation includes three relevant models:

**I2V**: This is an extension of I2V [10] (Section 2) to generate a user representation: At the inference phase, $i_{1:k}$ is modeled by the average of the context vectors $u_{i_{1:k}}$, and $i_{k+1}$ is the target vector $v_{i_{k+1}}$. Then, the similarity of $u_{i_{1:k}}$ and $v_{i_{k+1}}$ is computed at Eq. (3).

**NCF**: This is a recently published Neural Collaborative Filtering model [26] that fuses the Generalized Matrix Factorization method with Multilayer Perceptron for user-item recommendations. It was shown to outperform strong baselines like eALS [13] and BPR [6].

**AI2V**: The proposed AI2V model (Section 3).

### 4.4. Hyperparameters Configuration and Optimization

The evaluated models used embedding dimension $d = 100$. The number of negative samples that are drawn per a positive example is $|\mathcal{N}| = 8$. For I2V, we used a subsampling parameter of $3 \cdot 10^{-4}$. For AI2V, we further set the attention space dimension $d_a = 40$ and set $N = 1$ (Section 3.2).

AI2V and I2V were optimized using Adagrad [19] with learning rate 0.1 and minibatch size of 32, in accordance with [12]. NCF was optimized according to [13]. All models were optimized until a saturation in validation measures (~40 epochs). No overfitting was observed.

### 4.5. Results

Tables 2-4 present HR@5, HR@10, HR@20, MRR@5, MRR@10, MRR@20 for all combinations of models and datasets. Note that unlike the original work of NCF [13], we used a different preprocessed version of MovieLens 1M dataset, and thus the results are different from the ones reported in [13].

On both Yahoo! Music and MS datasets, AI2V significantly outperforms both I2V and NCF. On the Movielens dataset, AI2V is competitive and provides slightly better results than I2V, and one can conclude that attention mechanisms are preferable on this dataset as well.

Overall, the results in Tabs. 2-4 indicate that the AI2V method is useful for learning attentive user representations and is at least competitive with other I2V and NCF.

## 5. CONCLUSION

This work presents AI2V – a novel attentive version of I2V that employs a context-target attention mechanism. AI2V attends the historical context items consumed by the user w.r.t. a potential target item recommendation, and produces a neural attentive user representation. The effectiveness of AI2V is demonstrated on several datasets, where it is shown to outperform both I2V and NCF.